\documentclass[acmsmall]{nimeart}
\usepackage[table,xcdraw]{xcolor}
\usepackage[plain]{fancyref}
\setcopyright{cc}
\copyrightyear{2026}
\acmYear{2026}
\acmDOI{}
\acmConference[NIME '26]{International Conference on New Interfaces for Musical Expression}{June 23--26,
  2026}{London, UK}
\acmISBN{}

\begin{document}

\title{The Moving Drone: Negotiating Agency Between the Voice and the Virtual}

% Author information, leave this blank for the initial submission.
\author{Nithya Shikarpur}
\affiliation{%
  \institution{Massachusettes Institute of Technology}
  \city{Cambridge}
  \country{USA}
}

\author{Victor Arul}
\affiliation{%
  \institution{Harvard University}
  \city{Cambridge}
  \country{USA}
}

\author{Anna Huang}
\affiliation{%
  \institution{Massachusettes Institute of Technology}
  \city{Cambridge}
  \country{USA}
}


\maketitle

\begin{figure}[h]
    \centering
    \includegraphics[width=0.9\linewidth]{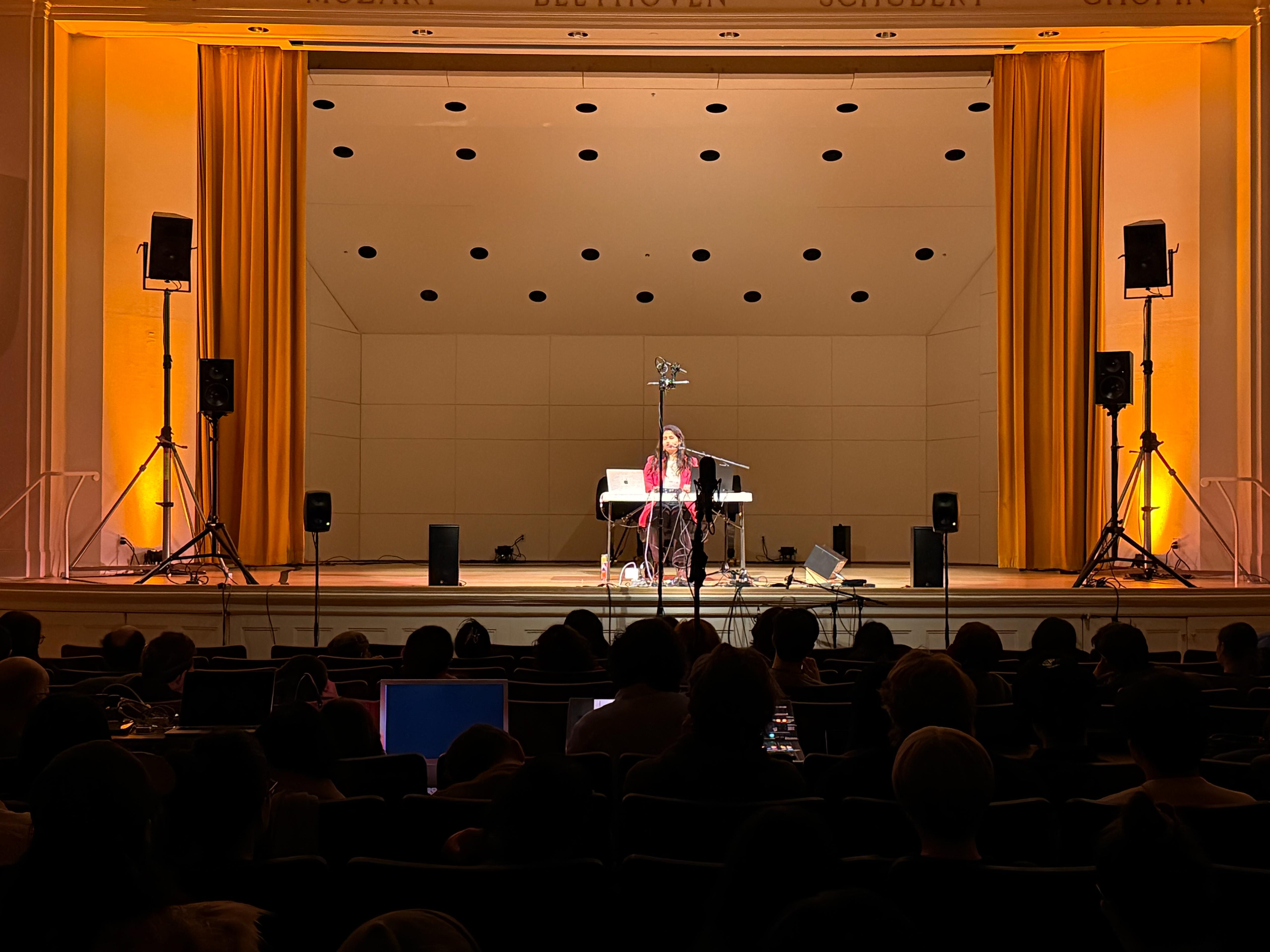}
    \caption{A still from the premiere of 'The Moving Drone' at Harvard University, featured in the Hydra concert series.\protect\footnotemark}
    \label{fig:performance-still}
\end{figure}
\footnotetext{The speaker set up in this figure is different, please refer \fref{fig:tech-sheet} for the NIME performance proposed set up.}

% Rather than chasing the high-fidelity realism that has sparked industry-wide anxiety over job replacement, this work intentionally utilizes "lo-fi" generative outputs. By requiring human interpretation and situational context, "The Moving Drone" positions AI not as a replacement, but as a responsive, co-creative partner rooted in established socio-cultural practice.

\section{Program Notes}
Melodic material in Hindustani music is presented in relation to a tonic, usually sustained by the \textit{tanpura}, a four-stringed drone instrument. Rooted in Hindustani music, `The Moving Drone' sets the traditionally static drone into motion that, throughout the performance, gains increasing agency transitioning from reactive to more proactive roles. The work employs four independent loopers in Max/MSP to function as `virtual' drones. They are populated cyclically in real-time as the vocalist improvises, creating an organic and evolving feedback loop between the voice and the virtual drone. This relationship further evolves melodically by pitch shifting the loops, which introduces a dimension of sudden, explicit movement. Then it changes timbrally, via the integration of GaMaDHaNi, a singer conditioned pitch-to-voice generative AI model to resynthesize looped audio. While current music AI approaches prioritize high-fidelity and realism of generated content which has sparked anxiety over job replacement for the music community, this work intentionally utilizes low-fidelity generative outputs, further necessitating human interpretation and situational context in order to be complete. `The Moving Drone' positions technology and generative AI within established socio-cultural musical practices, proposing a virtual drone as an active, responsive, and co-creative musical agent.

% what is a raag?
\section{Project Description}
\label{sec:description}

\begin{figure}
    \centering
    \includegraphics[width=0.9\linewidth, trim={5.5cm 2cm 5.5cm 2cm}, clip]{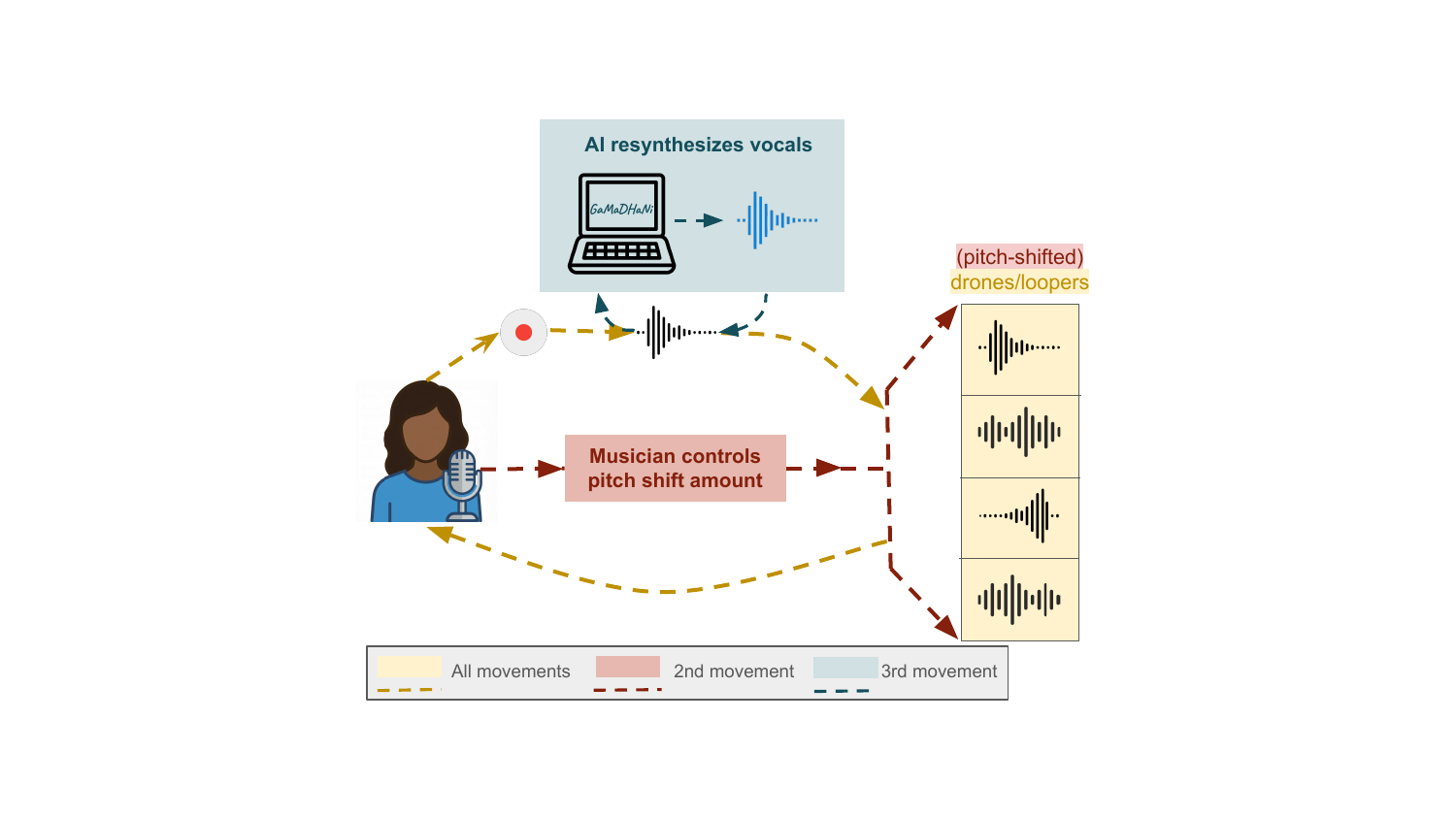}
    \caption{High-level system design for the performance across three movements.}
    \label{fig:system-design}
\end{figure}

Hindustani music, i.e. North Indian Classical music, is a modal tradition where melodic meaning of each note is derived from it's relationship with a fixed tonic \cite{bagchee1998nad}. This relationship is externalized by the \textit{tanpura}, a four-stringed drone that plays the tonic in addition to other fixed notes to lay the steady melodic foundation upon which the performer improvises. Through this performance, we explore giving increased agency to this conventionally passive drone and probe the negotiation between vocal improvisation and drone. 

In this work, the drone is reimagined as a virtual entity realized through four independent loopers that are cyclically populated by the performer's live improvisations, creating a real-time feedback loop. We conceptualize the drone's agency along two axes, pitch and timbre, each ranging from reactive (low agency) to proactive (high agency) \cite{thelle2021spire} \footnote{This is a work in progress, and we acknowledge that the full scale of the theoretical framework has not been utilized yet, but we hope to use this design space as a guide for future work.}.

Each looper, implemented in Max/MSP \cite{cycling}, has a fixed buffer length of three seconds, and individual controls on the amount of crossfade to prevent clicks as the loop repeats. The performer controls what is recorded into loops by means of a manually controlled record button which cyclically populates the loops. The high-level system layout is depicted in \fref{fig:system-design}. The performance consists of three movements: (1) organically evolving loops, (2) jumping loops, and (3) AI-resynthesized loops, each exploring different aspects of melody and timbre elaborated below. 

\subsection{Movement 1: Organically Evolving Loops}

The performance starts with the loops playing the tonic (\textit{S}) and the dominant (\textit{P}), similar to a traditional \textit{tanpura}. The vocalist improvises in raga \textit{Bihag}, a raga replete with feelings of love and happiness \cite{raag-bihag, ashwini-bhide}. Structurally, the emphasis on the notes \textit{S, G, P, N} (analogous to a Major 7th chord) provides a stable and fairly consonant harmonic landscape for the drone. As the vocalist sings, specific tones are recorded into the loops to create a symbiotic, ever-evolving virtual drone. In this movement, the drone’s agency remains limited, functioning reactively across both melodic and timbral dimensions.

\subsection{Movement 2: Jumping Loops}

This movement introduces a more explicit, `forced' movement along the melodic axis, transitioning the drone into a relatively more proactive role. While the movement begins with the standard tonic (\textit{S}) and fifth (\textit{P}), the system no longer simply records and loops the vocalist. Instead, movement is shown by pitch-shifting loops across four pre-determined presets, each corresponding to a distinct raga: \textit{Bihag}, \textit{Bhairavi}, \textit{Basant}, and \textit{Kafi}. These presets are designed so that most absolute pitches remain consistent across the four ragas, similar to ideas in melodic mode switching or \textit{grihabedham} \cite{bagchee1998nad} as seen in \fref{tab:movement2}. Specific notes function as `portals' or pivots that allow the performer to transition between these varied tonal landscapes and the raga characteristic phrases pose as anchors to help anchor oneself in the constantly changing raga space. In this phase, the system actively shapes the vocal improvisation rather than merely reflecting it.

\begin{table}[]
\begin{tabular}{@{}cc|cccccccccccc@{}}
\toprule
Raga              & \begin{tabular}[c]{@{}c@{}}Pitch shift\\ amount\end{tabular} & A\#                               & B & C                                 & C\# & D                                 & D\#                               & E                                 & F                                 & F\# & G & G\# & A                                 \\ \midrule
Bihag    & 0                                                                     & {\color[HTML]{CB0000} \textbf{S}} &            & R                                 &              & {\color[HTML]{CB0000} \textbf{G}} & {\color[HTML]{CB0000} \textbf{m}} & M                                 & {\color[HTML]{CB0000} \textbf{P}} &              & D          &              & {\color[HTML]{CB0000} \textbf{N}} \\
Bhairavi & -1                                                                    & r                                 &            & {\color[HTML]{CB0000} \textbf{g}} &              & m                                 &                                   & P                                 & {\color[HTML]{CB0000} \textbf{d}} &              & n          &              & {\color[HTML]{CB0000} \textbf{S}} \\
Basant   & 0                                                                     & {\color[HTML]{CB0000} \textbf{S}} & r          &                                   &              & G                                 & m                                 & {\color[HTML]{CB0000} \textbf{M}} & P                                 & d            &            &              & N                                 \\
Kafi     & +2                                                                    & d                                 &            & {\color[HTML]{CB0000} \textbf{n}} &              & {\color[HTML]{CB0000} \textbf{S}} &                                   & R                                 & {\color[HTML]{CB0000} \textbf{g}} &              & m          &              & {\color[HTML]{CB0000} \textbf{P}} \\ \bottomrule
\end{tabular}
\caption{Table showing the different ragas used during Movement 2. Notes are depicted using solfege notation in \cite{ganguli2018distributional} with A\# as the tonic. Notes highlighted in red are used as `portals' to jump into and out of each raga as the pitch shifting of loopers occurs.}
\label{tab:movement2}
\end{table}

\subsection{Movement 3: AI resynthesized loops}

In the final movement, the drone achieves timbral agency, mediated by GaMaDHaNi \cite{shikarpur2024hierarchical}, a hierarchical generative model trained on Hindustani vocal music. The model employs two stages: a pitch generator that defines the `melodic idea' and a spectrogram generator conditioned on a singer ID that realizes the melodic idea into sound. This is implemented by routing audio from Max/MSP to a dedicated GPU-powered laptop for real-time synthesis. Using GaMaDHaNi’s spectrogram generator, the input is transformed into another timbre before being fed into the loopers, with each looper assigned a pre-determined singer ID. 

Artistically, we choose to create a sense of chaos and tension in this section by leaning into the noisy and distorted quality of generated samples from the model, in part due to the 16 kHz sample rate and also the Griffin-Lim algorithm \cite{griffin1984signal} used to estimate phase. Additionally, raga \textit{Shree} chosen for improvisation, lends chromatic clusters that add to the harmonic dissonance in the drone. 

\section{Performer's Reflection: The Process of Negotiation}
\label{sec:negotiation}

The aesthetic impetus for this work was the transformation of the traditionally static drone into a dynamic, moving entity. Negotiating agency with this virtual drone necessitated a process of cognitive and technical adaptation, which I have distilled here to contextualize the design choices detailed in \fref{sec:description}.

First, the lack of a constant tonic from the tanpura, an element I had taken for granted, initially resulted in significant musical disorientation. Through practice, I developed an improvisation style that periodically revisited the tonic, ensuring that it was present in at least one of the four loops at all times. I also found myself relying on raga as a cognitive framework to find my footing in this new, changing sonic environment. Raga-characteristic phrases played the role of perceptual anchors for me, while the consonance or dissonance of notes in the raga guided the emotional intention for each movement. Furthermore, I recognized that populating the drone exclusively with my live vocal input constrained the system’s spectral width to the physical limits of my own vocal range. As a result, I adopted layers of octave-shifted loops to make the sound more full. The artificial voice felt uncanny marked by a quality that was both not-me and distinctly non-human. This tense and chaotic aesthetic was further amplified by the chromatic note clusters and austere nature of raga \textit{Shree}.

Exploring this interaction with the drone took me through a process of learning, understanding and unlearning that I hope to keep exploring further.

\section{Reflections: AI and Music}

The rapid advancement of text-to-music generative models has sparked strong backlash from music communities \cite{herington2025musicians, adam-neely}, exposing divergent motivation across, artistic, academic, and commercial entities \cite{ai-music-labels, mcpherson2019musical}. Previous work \cite{bindi2023ai} highlights the importance of situating AI application for music within cultural and artistic practices. However, the prevalence of standardized benchmarks \cite{kilgour2018fr, copet2023simple} often narrows the scope of evaluation, thus marginalizing the experimental, niche and culturally-informed approaches to music making and generation. Furthermore, Eurocentric bias-spanning datasets, model architectures, and task definitions continue to be a point of contention for artists from diverse musical backgrounds \cite{newman2023human, solak2025bias, shikarpur2024hierarchical, shikarpur2024exploratory}. 

This work adopts an intentional stance of grounding AI within the first author's practice of Hindustani music. By challenging the metrics prioritized by current evaluation benchmarks, such as high-fidelity and realism, we advocate for a more critical adoption of AI in music. We posit that music does not exist in a vacuum; it is a practice forged through centuries of culture and community. Acknowledging this lineage is essential when developing new musical technologies.

\section{Technical Notes}

\begin{figure}
    \centering
    \includegraphics[width=0.9\linewidth, trim={5cm 2cm 5cm 2cm}, clip]{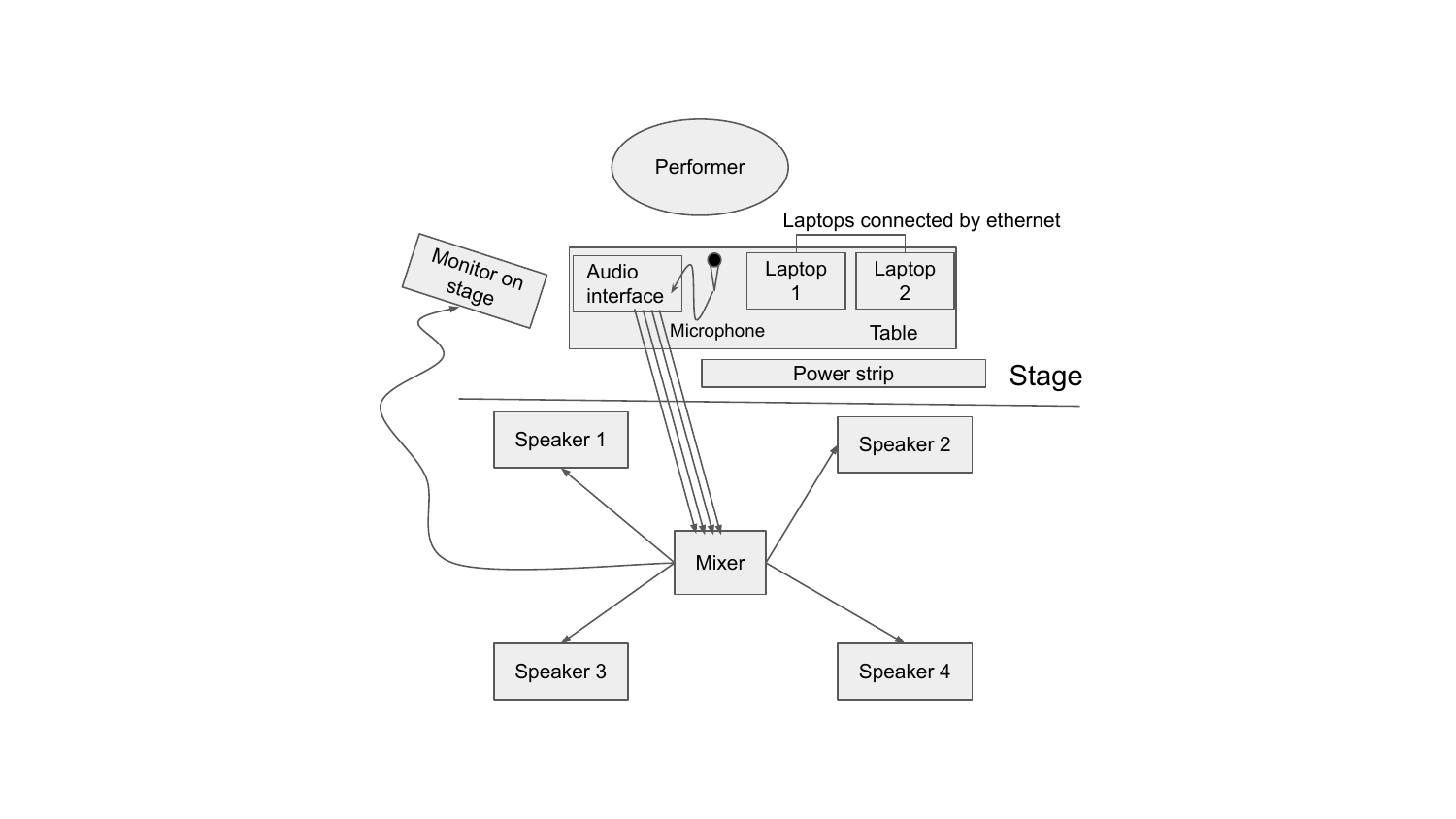}
    \caption{Technical Floor Plan}
    \label{fig:tech-sheet}
\end{figure}

This performance is a solo improvised set of around 13 minutes duration. Two laptops will be used, one to run Max/MSP and another laptop to run the model. The piece has a 4-channel output that can be reduced to stereo if required. See figure \ref{fig:tech-sheet} for a technical floor plan. We list the equipment required below:
\begin{itemize}
    \item Equipment provided by artists:
    \begin{itemize}
        \item Audio interface
        \item 2 Laptops
        \item Ethernet Cable
    \end{itemize}
    \item Equipment requested from organizers:
    \begin{itemize}
        \item Microphone along with a stand
        \item XLR for microphone
        \item 4 speakers with cables, each connected to 4 channel output from laptop (provided through Dante Virtual Soundcard)
        \item 1 Monitor on stage for the performer
        \item Table to place laptops during performance
        \item Power strip close to the table to charge laptops during performance
    \end{itemize}
\end{itemize}

\section{Media Links}
Media link: \url{https://youtu.be/3dJOzoxGx_c}

\section{Ethical Standards}
This is a solo performance and thus was not subject to the institute's ethical board review. This performance involves generative models trained on open source datasets \cite{srinivasamurthy2021saraga, gulati2016time, gulati2016phrase} comprising of around 120 hours of data. Although appropriate permissions were obtained from stakeholders, artists and labels, at the time of data collection, we acknowledge that the scope of generative modeling possibly fell outside of the original intent of researchers and artists. This discrepancy highlights the ongoing need for renewed ethical and licensing standards within the context of AI-driven musical co-creation. Additionally, we acknowledge the costly environmental impact of AI models and as a result strive to maintain small models that support both fast inference and also consume less power.

% % The acknowledgements section is optional and can be removed if not needed.
\begin{acks}
Nithya would like to thank Hans Tutschku for his invaluable guidance and feedback throughout the course of developing this performance. She also expresses deep gratitude to Weilu Ge and Manaswi Mishra who, through conversations, helped fuel and spark ideas for the performance. Most importantly she is grateful for all her music teachers who gave her the vocabulary and insight to develop these ideas.
\end{acks}

\bibliographystyle{ACM-Reference-Format}
\bibliography{sample-references}

@book{bagchee1998nad,
  title={Nad},
  author={Bagchee, Sandeep},
  year={1998},
  publisher={BPI Publishing}
}

@inproceedings{thelle2021spire,
  title={Spire muse: A virtual musical partner for creative brainstorming},
  author={Thelle, Notto JW and Pasquier, Philippe},
  booktitle={NIME 2021},
  year={2021},
  organization={PubPub}
}

@article{ganguli2018distributional,
  title={On the Distributional Representation of Ragas: Experiments with Allied Raga Pairs.},
  author={Ganguli, Kaustuv Kanti and Rao, Preeti},
  journal={Trans. Int. Soc. Music. Inf. Retr.},
  volume={1},
  number={1},
  pages={79--95},
  year={2018}
}

@misc{cycling,
	author = {Cycling '74},
	title = {Max},
	howpublished = {\url{https://cycling74.com/products/max}},
	year = {1997},
	note = {[Accessed 12-02-2026]},
}

@misc{raag-bihag,
    author = {Sabyasachi (Rahul) Bhattacharya},
	title = {Raga Bihag: An uplifting melody},
	howpublished = {\url{https://sarod.com.au/raga-bihag-an-uplifting-melody/}},
	year = {2014},
	note = {[Accessed 12-02-2026]},
}

@inproceedings{shikarpur2024hierarchical,
  title={Hierarchical generative modeling of melodic vocal contours in hindustani classical music},
  author={Shikarpur, Nithya and Dendukuri, Krishna Maneesha and Wu, Yusong and Caillon, Antoine and Huang, Cheng-Zhi Anna},
  booktitle={ISMIR},
  year={2024}
}

@article{griffin1984signal,
  title={Signal estimation from modified short-time Fourier transform},
  author={Griffin, Daniel and Lim, Jae},
  journal={IEEE Transactions on acoustics, speech, and signal processing},
  volume={32},
  number={2},
  pages={236--243},
  year={1984},
  publisher={IEEE}
}

@article{herington2025musicians,
  title={Musicians’ ethical concerns about AI: an interview study},
  author={Herington, Jonathan and Borasi, Raffaella and Guerrero, Benjamin J and Miller, David E and Koerner, Blaire and Han, Yu Jung and Borys, Zenon and Roberts, Rachel},
  journal={AI \& SOCIETY},
  pages={1--14},
  year={2025},
  publisher={Springer}
}

@misc{adam-neely,
    author = {Adam Neely},
	title = {Suno, AI Music, and the Bad Future},
	howpublished = {\url{https://www.youtube.com/watch?v=U8dcFhF0Dlk}},
	year = {2026},
	note = {[Accessed 12-02-2026]},
}

@misc{ai-music-labels,
    author = {Eamonn Forde},
	title = {Musicians are deeply concerned about AI. So why are the major labels embracing it?},
	howpublished = {\url{https://www.theguardian.com/music/2025/dec/16/musicians-are-deeply-concerned-about-ai-so-why-are-the-major-labels-embracing-it}},
	year = {2025},
	note = {[Accessed 12-02-2026]},
}

@incollection{mcpherson2019musical,
  title={Musical instruments for novices: Comparing NIME, HCI and crowdfunding approaches},
  author={McPherson, Andrew and Morreale, Fabio and Harrison, Jacob},
  booktitle={New directions in music and human-computer interaction},
  pages={179--212},
  year={2019},
  publisher={Springer}
}

@article{bindi2023ai,
  title={AI (r) evolution--where are we heading? Thoughts about the future of music and sound technologies in the era of deep learning},
  author={Bindi, Giovanni and Demerl{\'e}, Nils and Diaz, Rodrigo and Genova, David and Golvet, Ali{\'e}nor and Hayes, Ben and Huang, Jiawen and Liu, Lele and Martos, Vincent and Nabi, Sarah and others},
  journal={arXiv preprint arXiv:2310.18320},
  year={2023}
}

@article{kilgour2018fr,
  title={Fr$\backslash$'echet audio distance: A metric for evaluating music enhancement algorithms},
  author={Kilgour, Kevin and Zuluaga, Mauricio and Roblek, Dominik and Sharifi, Matthew},
  journal={arXiv preprint arXiv:1812.08466},
  year={2018}
}

@inproceedings{copet2023simple,
    title={Simple and Controllable Music Generation},
    author={Jade Copet and Felix Kreuk and Itai Gat and Tal Remez and David Kant and Gabriel Synnaeve and Yossi Adi and Alexandre Défossez},
    booktitle={Thirty-seventh Conference on Neural Information Processing Systems},
    year={2023},
}

@inproceedings{newman2023human,
  title={Human-AI Music Creation: Understanding the Perceptions and Experiences of Music Creators for Ethical and Productive Collaboration.},
  author={Newman, Michele and Morris, Lidia and Lee, Jin Ha},
  booktitle={ISMIR},
  pages={80--88},
  year={2023}
}

@article{solak2025bias,
  title={Bias beyond Borders: Global Inequalities in AI-Generated Music},
  author={Solak, Ahmet and Gr{\"o}tschla, Florian and Lanzend{\"o}rfer, Luca A and Wattenhofer, Roger},
  journal={arXiv preprint arXiv:2510.01963},
  year={2025}
}

@article{shikarpur2024exploratory,
  title={Exploratory Study Of Human-AI Interaction For Hindustani Music},
  author={Shikarpur, Nithya and Huang, Cheng-Zhi Anna},
  journal={NeurIPS Creative AI Track},
  year={2024}
}

@article{srinivasamurthy2021saraga,
  title={Saraga: Open datasets for research on indian art music},
  author={Srinivasamurthy, Ajay and Gulati, Sankalp and Caro Repetto, Rafael and Serra, Xavier},
  journal={Empirical Musicology Review 2021; 16 (1): 85-98.},
  year={2021},
  publisher={The Ohio State University Libraries}
}

@inproceedings{gulati2016time,
  title={Time-delayed melody surfaces for r{\=a}ga recognition},
  author={Gulati, Sankalp and Serr{\`a} Juli{\`a}, Joan and Ganguli, Kaustuv Kanti and Sent{\"u}rk, Sertan and Serra, Xavier},
  year={2016},
  booktitle={ISMIR}
}

@inproceedings{gulati2016phrase,
  title={Phrase-based r{\=a}ga recognition using vector space modeling},
  author={Gulati, Sankalp and Serra, Joan and Ishwar, Vignesh and Sent{\"u}rk, Sertan and Serra, Xavier},
  booktitle={2016 IEEE International Conference on Acoustics, Speech and Signal Processing (ICASSP)},
  pages={66--70},
  year={2016},
  organization={IEEE}
}

@misc{ashwini-bhide,
    key = {Raag Ki Tasveer - Bihag Family | Batiyan Daurawat | Dr. Ashwini Bhide Deshpande},
    title = {Raag Ki Tasveer - Bihag Family | Batiyan Daurawat | Dr. Ashwini Bhide Deshpande},
    author = {Ashwini Bhide Deshpande},
    note = {Accessed 04-30-2026},
    year = {2023},
    howpublished = {\url{https://www.youtube.com/watch?v=BptPdYSxRGs&list=PLEfmeUsduoEadg1CroTkD1pghArNZ7HkT&index=2}}
}

\end{document}